\documentclass[aps,prl,twocolumn,showpacs,groupedaddress]{revtex4-1}  
\usepackage{graphicx,color,dcolumn,bm,longtable,mathptmx,soul,float}
\usepackage{amsfonts,amsmath,amssymb}

\newcommand{\kk}{{\bf k}}  
\begin{document}

\title{Tunability of Magnetic Anisotropy of Co on Two-Dimensional Materials by Tetrahedral Bonding}
\author {D. Odkhuu$^1$}
\email{odkhuu@inu.ac.kr}
\author{P. Taivansaikhan$^{1}$ }
\author{N. Park$^{2}$}
\author{S. C. Hong$^3$}
\author{S. H. Rhim$^3$}
\email{sonny@ulsan.ac.kr}
\affiliation{$^1$Department of Physics, Incheon National University, Incheon 22012, South Korea \\
$^2$Department of Physics, Ulsan National Institute of Science and Technology, Ulsan 689-798, South Korea \\
$^3$Department of Physics, University of Ulsan and EHSRC, Ulsan 44610, South Korea }

\begin{abstract}
Pairing of $\pi$ electronic state structures with functional or metallic atoms makes them possible to engineer
physical and chemical properties.
Herein, we predict the reorientation of magnetization of Co on {\em hexagonal} BN ($h$-BN)
and graphene multilayers. The driving mechanism is the formation of the tetrahedral bonding
between $sp^3$ and $d$ orbitals at the interface.
More specifically, the intrinsic $\pi$-bonding of $h$-BN and graphene is transformed to $sp^3$
as a result of strong hybridization with metallic $d_{z^2}$ orbital.
The different features of these two tetrahedral bondings, $sp^2$ and $sp^3$, are well manifested
in charge density and density of states in the vicinity of the interface,
along with associated band structure near the $\bar{K}$ valley.
Our findings provide a novel
approach to tailoring magnetism by means of degree of the interlayer hybrid bonds in 2D layered materials.
\end{abstract}

\pacs{75.70.Cn, 73.20.Hb, 73.20.-r, 77.55.Nv}

\maketitle
\subsection {I. INTRODUCTION}
Two-dimensional (2D) materials, down to one or a few atomic thick, have been still active in modern material physics, including spintronics \cite{bai88,zut04}, spin-orbitronics \cite{hoff15}, and spin Hall effects \cite{hir99,hoff13},
owing to the intriguing properties associated with their unique atomic and electronic structures.
In particular, intense research efforts continue to seek possibilities
for engineering the physical and chemical properties of mono and multilayer graphene \cite{nov04,nov05,li09}
to fulfill the technological prerequisites,
where attempts by decorating with functional species or/and metals are prominent \cite{gio08,od13}.
On the other hand, the presence of graphene can also substantially alter the electronic and magnetic properties of
transition metal (TM) atoms, which depends on the degree of hybridization between the metal {\em d}
and graphene $\pi$ orbitals \cite{kra09}.
In these context, the exploration in other archetypal 2D structures such as
\emph{hexagonal} boron nitride (\emph{h}-BN), which is a structural analogue of graphene
with broken sublattice symmetry,
emerges one of central interests from both fundamental and technological stand points.

In spintronics,
perpendicular magnetic anisotropy (PMA) is of crucial significance,
where its tunability is essential to enhance the performance.
Furthermore, combination with 2D materials is yet to explore
the possibilities for manipulating the magnetism and magnetic anisotropy of TM atoms on graphene \cite{xia09,don13,don14,bai15}.
For example, unexpectedly large PMA up to an order of 100 meV
was predicted in cobalt dimer-benzene pairs \cite{xia09}.
Nevertheless, subsequent experiments have shown that
the individual Co atoms adsorbed onto graphene on a Pt(111) exhibit in-plane magnetic anisotropy \cite{don13}.
Interestingly, it is further identified that the preferable magnetization direction of Co adatoms on graphene depends on the underlying metallic substrate:
magnetization is
perpendicular in graphene/Ru(0001) and in-plane in graphene/Ir(111) \cite{don14},
where different graphene/metal interactions are responsible,
i.e. chemisorption in graphene/Ru, physisorption in graphene/Ir \cite{don14}, and graphene/Pt \cite{don13}.
In recent studies, through C \emph{p}$_z$-Co \emph{d}$_{z^2}$ hybridization,
the presence of fullerene molecules necessitates magnetization reorientation of
the underlying Co films in-plane to perpendicular \cite{bai15}.

In addition to these remarkable findings, the
promising alternative for tailoring the magnetic anisotropy of
TM atoms seemingly resides in the use of even stronger bonding
features between the tetrahedral $sp^3$ and metallic $d$ orbitals.
In this article, the reorientation of magnetization of Co is predicted
on $h$-BN via $sp^2-sp^3$ transition.
The bond transition includes hybridization with metallic $d$ orbitals,
where the $sp^3$-\emph{d}$_{z^2}$ hybridization is crucial for the magnetization reversal.
Moreover, the bond transition also accompanies the sign change in Berry curvature,
which may invites experiments to verify using the spin Hall\cite{guo:08,hir99,rmp:SHE}
and the inverse spin Hall effect\cite{wunderlich09,wunderlich10}.
Further investigations demonstrate
the reorientation of magnetization in similar phase transition
involving carbon $sp^3$-bonds from graphene layers \cite{raj13}.

\subsection {II. COMPUTATIONAL METHOD}
Density functional theory (DFT) calculations are performed
using Vienna \emph{ab initio} simulation package (VASP) \cite{paw94,kre93,kre96}.
The exchange-correlation interactions are treated
by the generalized gradient approximation formulated by Perdew, Burke, and Ernzerhof (PBE) \cite{per96}.
The long-range dispersion corrections for the interlayer interaction were taken into
account within the semi-empirical DFT-D2 approach suggested by Grimm \cite{grim06}.
We used $2\times2$ graphene and $h$-BN cells in the lateral
2D-lattice for all the calculations.
More details of our model geometries are provided in the following section.
We use an energy cutoff of 400 eV, a 21$\times$21$\times$1 \emph{k}-mesh,
and relaxation with force criteria of 10$^{-2}$ eV/\AA.
The spin-orbit coupling (SOC) is included in a second-variational way
employing scalar-relativistic calculations of the valence states \cite{koe77}.
Magnetic anisotropy energy (MAE) is calculated from the total energy differences between
when the magnetization is along the \emph{xy} plane (\emph{E}$^\|$) and along the \emph{z} axis (\emph{E}$^\bot$),
MAE $=$ \emph{E}$^\|$ -- \emph{E}$^\bot$.
A dense \emph{k} points of 41$\times$41$\times$1 is used in non-collinear calculations,
which is sufficient to get reliable results of MAE.

\begin{figure}[t]
  \centering
 \includegraphics[width=1\columnwidth]{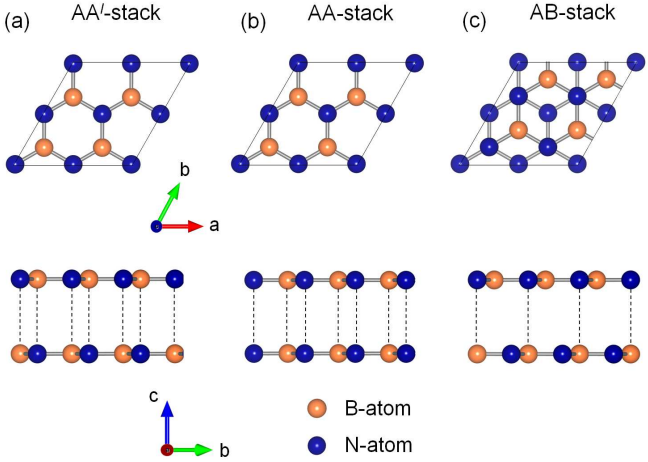}
   \caption{\label{fig1} Top- and sideview of three kinds of $h$-BN stacking:
(a) AA', (b) AA, and (c) AB-stacking,
where orange and blue sphere denote B and N atoms, respectively.
}
\end{figure}

\subsection {III. RESULTS AND DISCUSSION}
Prior to explore the Co$|$$h$-BN heterostructures,
we first examine the stacking and binding affinity of $h$-BN layers.
Top and side views of three distinctive stacking configurations of
$h$-BN bilayers are shown in Fig.~\ref{fig1}: (a) AA$^\prime$-stacking, (b) AA-stacking,
and (c) AB-staking (often referred to Bernal-type).
In the AA$^\prime$-stacking configuration,
each atomic column along the $c$-axis
consists of alternating B and N atoms.
In the AA-stacking configuration,
two $h$-BN layers are laterally superimposed.
In the AB-stacking configuration,
while B (N) atoms of the top layer
sit right above N (B) of the bottom layer,
the other atoms are above the hollow-site.
With the same pattern of the AB-stacking,
one may consider two more configurations,
i.e., A$^\prime$B and AB$^\prime$ \cite{rib11}, where
B (N) atoms of the top layer
sit right above B (N) of the bottom layer.
These two stacking structures have been excluded in the present study
owing to higher energies than AB-stacked $h$-BN
in previous \emph{ab initio} calculations \cite{rib11}.
Optimized in-plane lattice $a$, the
interlayer distance $d$, and the binding energy $E_b$ are listed in Table I
for the AA$^\prime$, AA, and AB stacked $h$-BN bilayers.
Here, 
$E_b=E(\textrm{BN bilayer})-2E(\textrm{BN monolayer})$,
where $E(\textrm{BN bilayer})$ and $E(\textrm{BN monolayer})$
are the total energies of bilayer and monolayer $h$-BN, respectively.
Overall, our structural results agree well with experimental \cite{han08,war10} and other theoretical studies \cite{rib11,liu03}.
The AB stacking is more stable by $\sim$3 meV
than the bulk-like AA$^\prime$ stacking,
in agreement with the previous calculations (4 meV) \cite{rib11}.
Experiments have also reported that both the AA$^\prime$ and AB stackings
can exist \cite{war10}, as their energy difference is quite small in our calculations.

   \begin{table} [t]
  \caption{\label{tab:1} Optimized in-plane lattice \emph{a} (\AA) and
    interlayer distance \emph{d} (\AA), and binding energy $E_b$ (eV per $1\times1$ unit cell)
        of $h$-BN bilayer for the different stacking configurations shown in Fig.~\ref{fig1}.}
  \begin{ruledtabular}
\begin{tabular}{ccccccccc}
 & \emph{a} & \emph{d} & $E_b$ \\
\hline
AA$^\prime$ & 2.513 & 3.145 & --0.138 \\
AA & 2.500 & 3.250 & --0.082   \\
AB & 2.513 & 3.152 & --0.141   \\
\end{tabular}
\end{ruledtabular}
\end{table}

\begin{figure}[b]
  \centering
 \includegraphics[width=1\columnwidth]{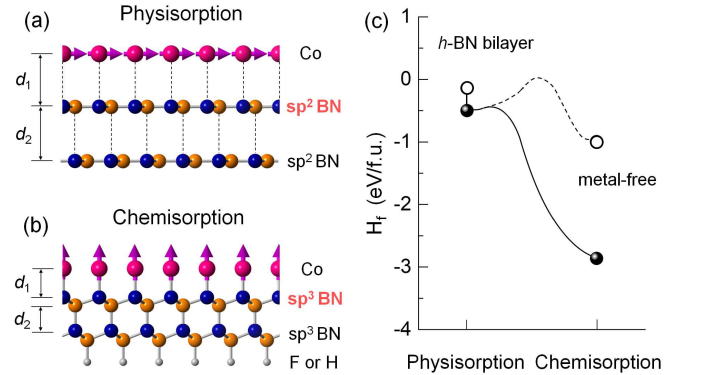}
   \caption{\label{fig2} Side views of the optimized atomic structures of a Co monolayer on the
  (a) $sp^2$ and (b) $sp^3$ BN bilayers.
  Red, blue, orange, and gray spheres represent the Co, N, B, and F atom, respectively.
  The arrows through the Co atoms denote the direction of Co magnetic moments.
  (c) The formation energy $H_f$ of the physisorbed and chemisorbed Co$|$BN.
  The case of metal-free chemisorbed BN bilayer is shown in open circle.
Total energy of Co adatoms on functional-free \emph{h}-BN is taken as reference energy.}
\end{figure}

The structural stability of Co$|$$h$-BN, upon functionalization of the bottom surface, is then explored, where we adopt only the AB-stacked $h$-BN bilayer,
whose structures are shown in Figs.~\ref{fig1}(a) and ~(b).
In Fig.~\ref{fig1}(a), Co(0001) monolayer is weakly physisorbed onto \emph{h}-BN bilayer with AB-stacking,
where vacuum region separating the periodic slabs is taken no less than 15 {\AA} thick.
The most preferred adsorption site of the Co adatoms on \emph{h}-BN is N-top site,
where the Co-BN interlayer distance, denoted as \emph{d}$_1$, is 3.32 $\textrm{\AA}$ (Table II).
In usual experiments,
functional atoms in gas phase can cover up to half the surface of graphene in a uniform pattern \cite{luo09,bal10}.
As such, fluorine atoms preferably bind with B
at the bottom surface of \emph{h}-BN bilayer \cite{zha11} as shown in Fig.~\ref{fig2}(b).
Consequently, the latter structure, Co$|$\emph{h}-BN$|$F,
is energetically favored than the former, fluorine-free Co$|$\emph{h}-BN,
by $\sim$1.8 eV/Co, from formation energy, $H_f$ \cite{od13}, as shown in Fig.~\ref{fig2}(c),
where $H_f=\left[E({\rm Co/BN/F})-E({\rm Co})-E({\rm BN})-E({\rm F})\right]/N$;
$E({\rm Co/BN/F})$, $E({\rm Co})$, $E({\rm BN})$, and $E({\rm F})$
are the total energies of Co/BN/F, $h$-BN bilayer,
and F atom in F$_2$-gas phase, respectively;
$N$ is the number of 1$\times$1 unit cell of $h$-BN in 2D lattice.

\begin{figure} [t]
  \centering
 \includegraphics[width=1\columnwidth]{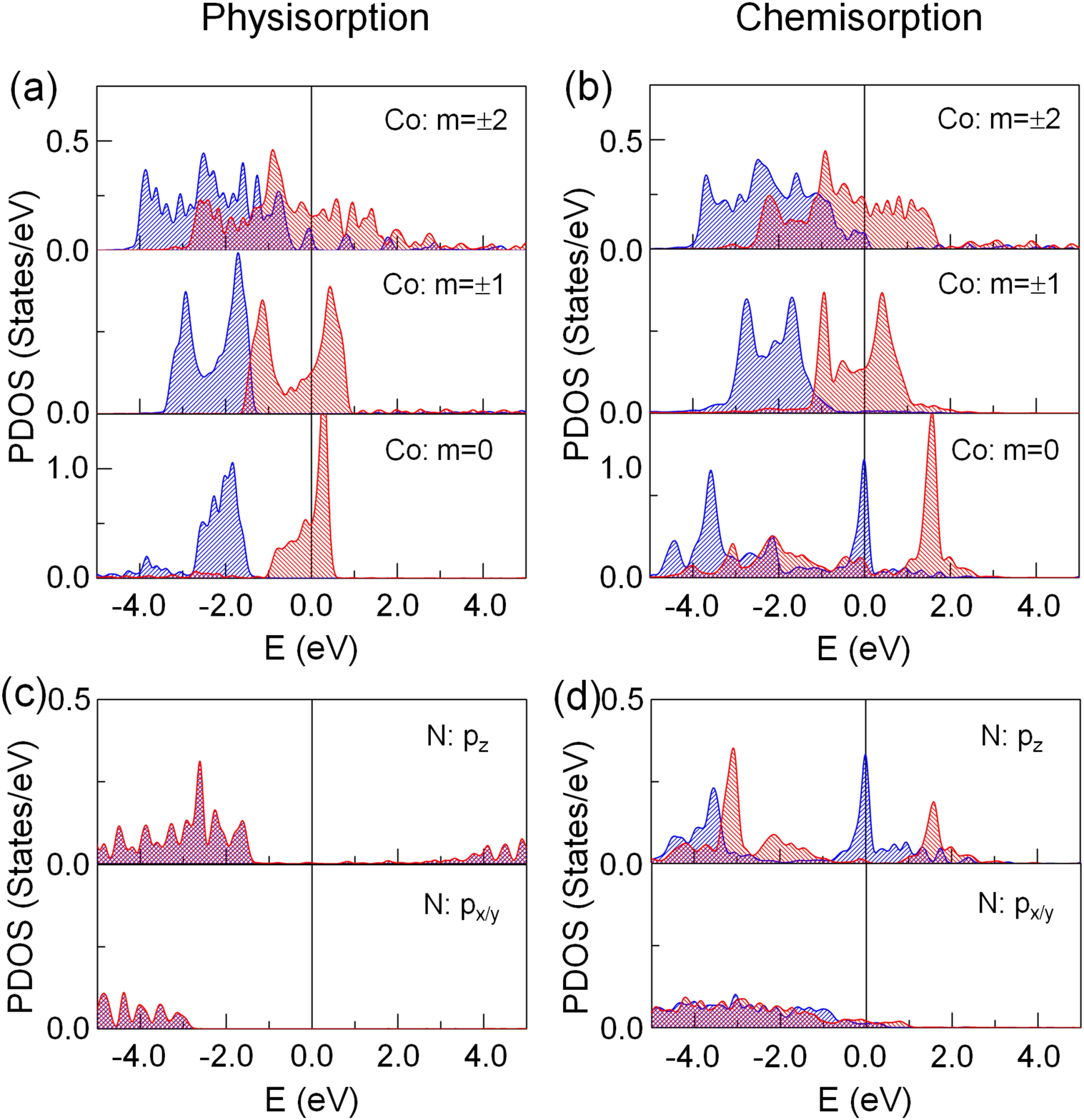}
 \caption{\label{fig3} The \emph{d}-orbital PDOS of the Co atom (a) physisorbed and
   (b) chemisorbed on BN bilayer.
   (c) and (d) The corresponding \emph{p}-orbital PDOS of the N atom underlying the Co atom.
   The blue and red areas denote the spin-up and spin-down states, respectively. The Fermi level is set to zero energy.}
 \end{figure}

Upon functionalization,
\emph{h}-BN layers either retain the van der Waals distance [Fig.~\ref{fig2}(a)]
or form covalent interlayer bonds [Fig.~\ref{fig2}(b)].
The optimized interlayer distance between the BN layers, denoted as \emph{d}$_2$,
of the former and the latter structures are 3.24 and 1.56 $\textrm{\AA}$, respectively (Table~\ref{tab:1}).
The reduced value of the latter characterizes $sp^3$ bond in {\em cubic} BN structure (\emph{c}-BN).
Remarkably, the Co$|$$sp^3$-BN structure is more favored
than the Co$|sp^2$-BN by $\sim$1.5 eV/Co, as shown in Fig.~\ref{fig2}(c).
Other studies have reported similar features
for functionalized (either hydrogenation or fluorination) graphene layers on metal surfaces \cite{od13,raj13}.
More importantly, the authors identified that
the energy barrier from graphene layers to \emph{sp}$^3$-bonded carbon transition is negligibly small.
Similarly, the transition in two-side fluorinated BN-layers demands no energy barrier \cite{zha11},
which is expected in the present system.

\begin{table} [b]
  \caption{\label{tab:2} Optimized in-plane lattice \emph{a} (\AA),
    interlayer distance \emph{d}$_1$ between the Co and the top BN layer
    and \emph{d}$_2$ between the BN layers (\AA),
    and spin magnetic moment $\mu_s$ ($\mu_B$) of the bonded Co and N atoms
    for the physisorbed and chemisorbed Co$|$BN.
    Orbital anisotropy $\Delta \mu_{\textrm{o}}$ ($\times 10^{-2} \mu_B$) and MAE (meV) are also listed.}
  \begin{ruledtabular}
\begin{tabular}{ccccccccc}
 & \emph{a} & \emph{d}$_1$ & \emph{d}$_2$ & $\mu_s^{\textrm{Co}}$ & $\mu_s^{\textrm{N}}$ & $\Delta \mu_\textrm{o}^{\textrm{Co}}$ & MAE \\
\hline
Physisorption & 2.49 & 3.32 & 3.24 & 1.98 & 0.00 & --0.79 & --0.62 \\
Chemisorption & 2.56 & 1.94 & 1.56 & 1.95 & 0.05 & 0.82 & 0.76 \\
\end{tabular}
\end{ruledtabular}
\end{table}

\begin{figure} [b]
  \centering
 \includegraphics[width=1\columnwidth]{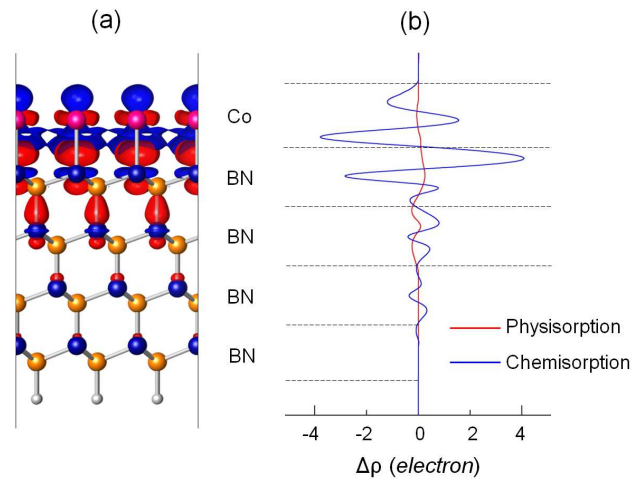}
 \caption{\label{fig4} (a) Isosurface plot of the charge density difference $\Delta \rho$,
   where red (blue) corresponds to charge accumulation (depletion)
   in unit of $5\times10^{-2}$ $e$/bohr$^3$.
(b) The planar average of the $\Delta \rho$ ($e$) along the \emph{z} axis
   in the case of Co/BN with $sp^3$-bonding (chemisorbed).
  The notation of the atomic symbol is the same as used in Fig.~\ref{fig1}.
  In (b), the planar average of the $\Delta \rho$ for the physisorption is also shown in red line.}
\end{figure}
From Fig.~\ref{fig2}(c), the presence of metal atom is essential
for the favorable formation of the interlayer bonds evidenced in $H_f$.
Here, the half-filled \emph{d} states of metal are served as a passivation of the otherwise unstable \emph{sp}$^3$ dangling bonds.
Noteworthy, the strong hybridization between the N-$p_z$ and Co-\emph{d}$_{z^2}$ orbitals is evident \cite{od13,raj13}.
Moreover, such \emph{sp}$^3$ bonded \emph{c}-BN structure turns out thermodynamically and structurally stable with the BN-thickness up to twenty atomic layers ($\sim$4nm).
This again reveals an important role of the metal atom in the interlayer formation.

To better appreciate the significance of the $sp^3$-$d_{z^2}$ hybridization,
partial density of states (PDOS) of the bonded Co and N atoms are shown in Figs.~\ref{fig3}(a)--(d)
for the physisorbed and chemisorbed Co$|$BN, respectively.
PDOS of the free-standing Co and \emph{h}-BN monolayers remain almost unchanged in the physisorption.
The two doublets, \emph{d}$_{xy/x^2-y^2}$ ($m=\pm2$) and \emph{d}$_{xz/yz}$ ($m=\pm1$) states, retain during the $sp^2-sp^3$ transition,
where $m$ is the magnetic quantum number.
In contrast, the chemisorption alters Co-\emph{d}$_{z^2}$ and N-\emph{p}$_z$ states,
where common peak in PDOS, owing to hybridization, is apparent.
Notably, the N-\emph{p}$_z$ states into the majority- and minority-spin subbands are split
by such strong hybridization,
which in turn leads to a large majority-spin peak at the Fermi level ($E_F$).

The split N-$p_z$ state in $E_F$ exhibits the feature of the  metal-induced gap states (MIGS),
which penetrates two- or three-layer-thick ($\sim$0.5nm) into the \emph{c}-BN layers.
Fig.~\ref{fig3} analyzes the electron density profile, 
$\Delta\rho=\rho(\rm{Co|BN})-\rho(\rm{Co})-\rho(\rm{BN})$,
where the MIGS is clearly manifested; 
the larger electronegativity (3.04) of the N than the Co (1.91) drives
the spin-polarized charge transfer from the metal atom.
On the other hand, the strong hybridization between the Co-$d$ and N-$p$ states
causes charge redistribution in the metal layer.
From the integration of the occupied Co-PDOS,
the \emph{d$_{z^2}$} orbital in the majority-spin loses 0.09$e$ under the formation of Co$|sp^3$-BN
while the \emph{d$_{xy/x^2-y^2}$} in the majority-spin and \emph{d$_{z^2}$} in the minority-spin gain 0.02 and 0.03$e$, respectively.

Remarkably, the $sp^2-sp^3$ transition accompanies reorientation of magnetization.
Calculated MAE and orbital moment anisotropy,
$\Delta \mu_{\textrm{o}}$ $=$ $\mu_{\textrm{o}}^\bot - \mu_{\textrm{o}}^\|$,
are listed in Table~\ref{tab:1} for the physisorbed and chemisorbed Co$|$BN, respectively.
The sign change for both MAE and $\Delta \mu_{\textrm{o}}$ emerges from the physisorption (--0.62 meV and --0.79$\times10^{-2} \mu_B$) to the chemisorption (0.76 meV and 0.82$\times10^{-2} \mu_B$),
which indicates that the Co magnetization reorients from in-plane to perpendicular direction
during the $sp^2-sp^3$ transition.
Moreover, these MAE values are fairly reproduced by the force method \cite{wei85}
from the \emph{k}-resolved MAE
according to $\text{MAE}(k)\approx\sum_{n\in{occ}}[\varepsilon(n,k)^\|-\varepsilon(n,k)^\bot]$ in 2D Brillouin zone (BZ),
where $\varepsilon(n,k)^\|$ and $\varepsilon(n,k)^\bot$
are eigenvalues of occupied states for in-plane and perpendicular magnetization, respectively.
Summing over \emph{k} points,
we find the MAE values of --0.59 and 0.76 meV for the Co$|sp^2$-BN and Co$|sp^3$-BN,
respectively, and both of them mainly arise from the $\text{MAE}(k)$ at/around the $\overline{\textrm{K}}$ point.

\begin{figure} [t]
  \centering
 \includegraphics[width=1\columnwidth]{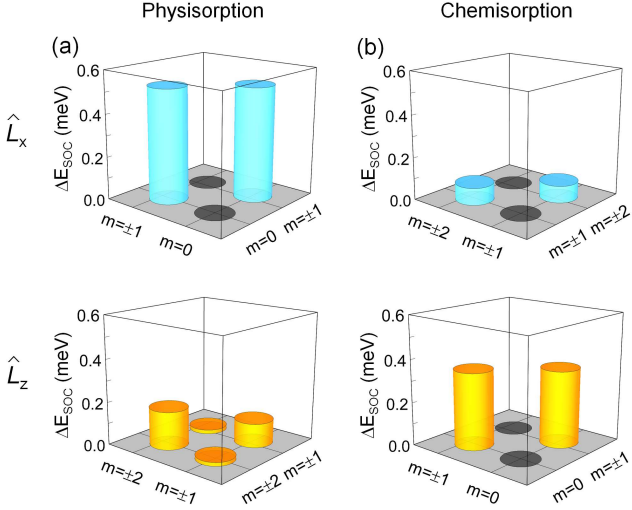}
  \caption{\label{fig5}Difference of \emph{d}-orbital projected SOC energies, $\Delta E_{\textrm{soc}}$, between in- and out-of-plane
  magnetization orientation of the Co atom (b) physisorbed and (c) chemisorbed on bilayer BN.
  Positive and negative contributions are denoted by orange and blue bars, respectively.}
\end{figure}

To elucidate the origin of magnetization reversal, we show
the $d$-orbital-projected contributions to the difference in SOC energies for in- and out-of-plane magnetization orientation,
$\Delta E_{\textrm{soc}}$ $=$ $E_\textrm{soc}^\|$ -- $E_\textrm{soc}^\bot$,
in Figs.~\ref{fig5}(a) and (b) for the physisorbed and chemisorbed Co$|$BN, respectively.
Here, $E_{\textrm{soc}} = <\frac{\hbar^2}{2m^2c^2}\frac{1}{r}\frac{dV}{dr} \textbf{\emph{L}}\cdot\textbf{\emph{S}}>$,
where \emph{V}(\emph{r}) is the spherical part of the effective potential within the PAW sphere, and \textbf{\emph{L}} and \textbf{\emph{S}} are
the orbital and spin angular momentum operators, respectively.
The expectation value of $E_{\textrm{soc}}$ is twice the actual value of the total energy correction to the second-order in SOC, i.e., MAE $\approx$ 1/2$\Delta E_{\textrm{soc}}$ \cite{ant}.
Our test calculations indicate that the second-order perturbation theory is a reasonable approximation as the total MAE overall agree within a few percent accuracy
with those obtained from the atom and orbital projected calculations.
The other 50\% of the SOC energy translates into the crystal-field energy and the formation of the unquenched orbital moment \cite{sko}.
It is obvious as seen in the top panel in Fig.~\ref{fig5}(a) that the negative MAE of the Co/$sp^2$-BN is predominated by the $ \langle xz,yz_  |\hat{L}_{x}|z^2 \rangle $,
where $\hat{L}_{x(z)}$ is the orbital angular momentum operator of the $x$ ($z$) component,
which serves as negative (positive) contribution to MAE \cite{wang93}.
Two positive contributions, involving the $m=\pm1$ and $m=\pm2$ states,
are rather small [bottom panel in Fig.~\ref{fig5}(a)].
Interestingly, as shown in Fig.~\ref{fig5}(b), these negative and positive contributions are
reversed in sign under the $sp^2-sp^3$ transition.

\begin{figure} [t]
  \centering
 \includegraphics[width=1\columnwidth]{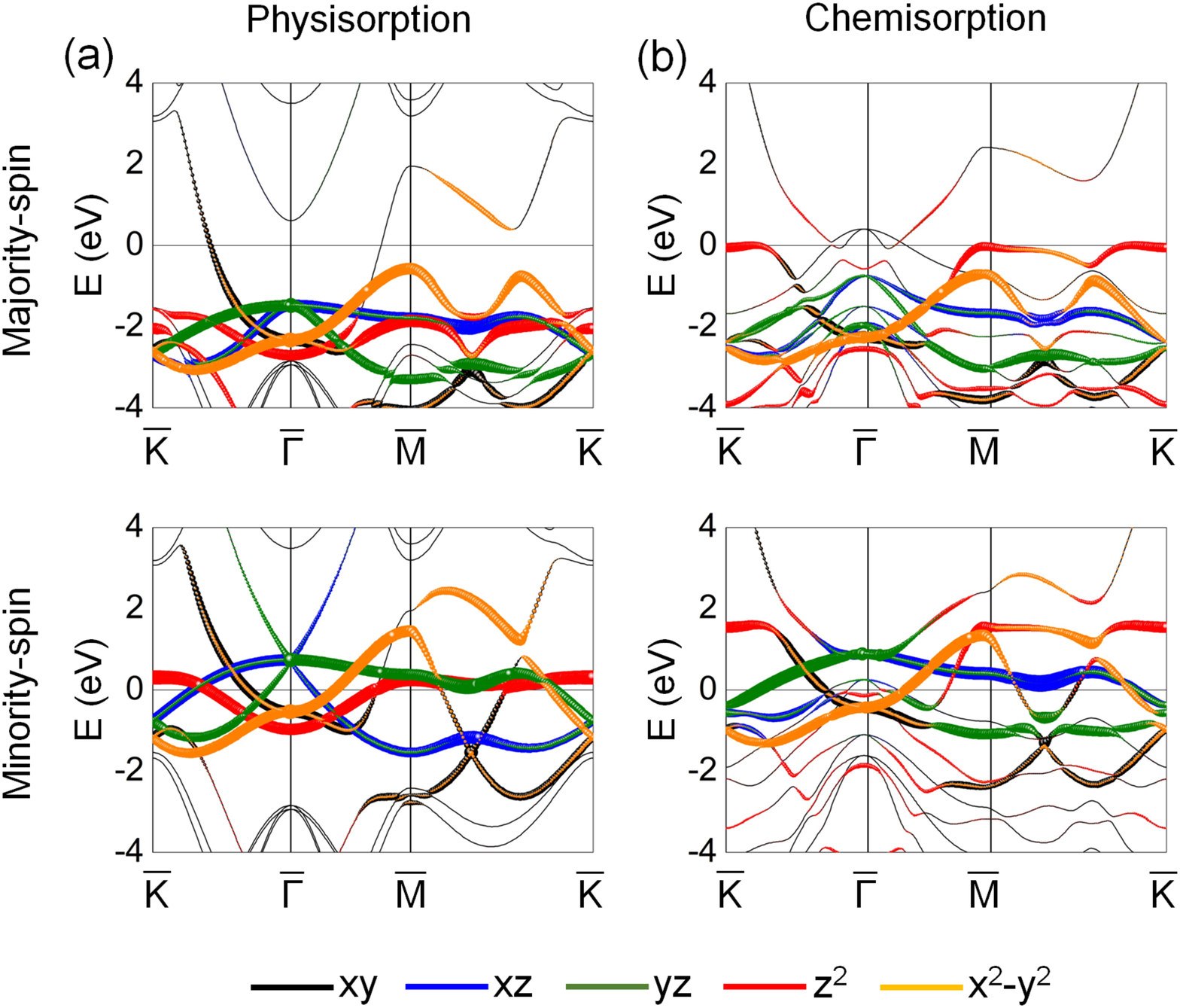}
  \caption{ \label{fig6}Energy- and \emph{k}-resolved distribution of the orbital character of the (a) and (b) majority- and (c) and (d) minority-spin
  bands of the physisorbed and chemisorbed Co/BN, respectively.
  The symbols superimposed over the band lines with black, orange, green, red, and blue colors
  represent the \emph{d}$_{xy}$, \emph{d}$_{xz}$, \emph{d}$_{yz}$, \emph{d}$_{z^2}$, and \emph{d}$_{x^2-y^2}$
  orbitals of the Co adatom, where the size of the symbols is proportional to their weights.
  The Fermi level is set to zero in energy.}
\end{figure}

\begin{figure} [t]
  \centering
 \includegraphics[width=1\columnwidth]{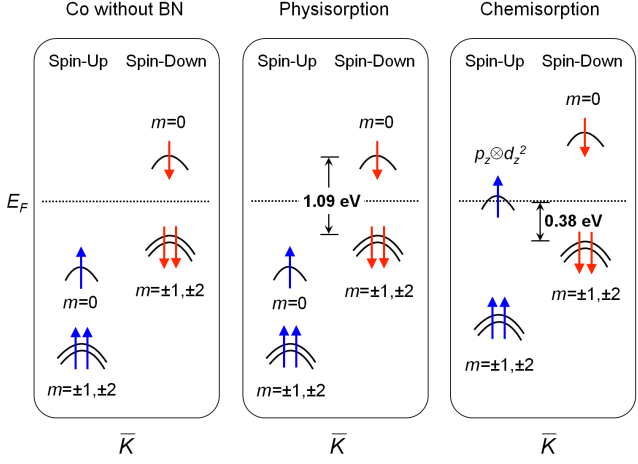}
  \caption{\label{fig7} Schematic diagram of the interlayer bond splitting of the
Co \emph{d}- and N \emph{p}-orbital levels at the $\overline{\textrm{K}}$ point in the 2D BZ.
The metal \emph{d}-orbital states are not much perturbed upon the physisorption with \emph{h}-BN (middle panel). The interlayer bonds at the Co$|sp^3$-BN interface splits the Co $d_{z^2}$ and N $p_z$ into the hybrid bonding states across the Fermi level ($E_F$),
denoted as $p_z\otimes d_{z^2}$ in the right panel. The upward red and downward blue arrows represent the majority- and minority-spin states, respectively.}
\end{figure}

In order to understand more,
we show energy- and $k$-resolved
distribution of the Co \emph{d}-orbital character of the majority- and minority-spin bands along the high-symmetry $\overline{\textrm{M}}\overline{\Gamma}\overline{\textrm{K}}\overline{\textrm{M}}$ line in Figs.~\ref{fig6}(a) and (b) for the physisorbed and chemisorbed Co$|$BN, respectively.
From this, in Fig.~\ref{fig7}, we sketch a schematic diagram
of Co $d$-orbital energy levels at the $\overline{\textrm{K}}$.
For reference, the same for the free-standing Co(0001) monolayer is also compared.
In the free-standing Co, without BN,
Co $d$ states are split into a singlet ($m=0$) and two doublets ($m=\pm1,\pm2$).
In the physisorption, while Co $d$ orbitals hybridize weakly with BN via $sp^2$ bonding,
energy levels do not change very much.
The SOC split energy between the occupied and unoccupied states near $E_F$
is $\varepsilon_{xz,yz}(\downarrow) - \varepsilon_{z^2}(\downarrow)\sim$1.09 eV,
which occurs in the minority-spin channel ($\downarrow$).
We simply ignore the $\uparrow \uparrow $- and $\uparrow \downarrow $-channel couplings as the majority-spin states are completely filled.
On the other hand, in the chemisorbed case,
as discussed earlier,
the strong hybridization at the interface
splits the Co $d_{z^2}$ and N $p_z$ into some hybrid bonding states,
denoted as $p_z\otimes d_{z^2}$ [right panel in Fig.~\ref{fig7}].
In particular, the hybrid $p_z\otimes d_{z^2}$ in the majority-spin channel ($\uparrow$)
contributes to the MIGS at $E_F$,
which in turn modifies SOC significantly.
Thus, from the energy level analysis along with the second-order perturbation theory of SOC matrix \cite{wang93},
the sign reversal in SOC matrix under the $sp^2-sp^3$ transition
mainly originates from two factors:
(i) $\varepsilon_{xz,yz}(\downarrow) - \varepsilon_{z^2}(\downarrow)\sim$ 1.94 eV,
which weakens the negative MAE in the magnitude,
(ii) $\varepsilon_{z^2}(\uparrow) - \varepsilon_{xz,yz}(\downarrow)\sim$ 0.38 eV,
which provides an additional strong contribution to PMA
through $ \langle z^2,\uparrow |\hat{L}_{z}| xz,yz,\downarrow \rangle $.

\begin{table} [b]
\caption{\label{tab:2} The formation energy $H_f$ (eV/Co), optimized in-plane lattice \emph{a} (\AA), interlayer distance \emph{d}$_1$ between the Co and the top carbon layer and \emph{d}$_2$ between the carbon layers (\AA),
spin magnetic moment ($\mu_B$) and orbital moment difference $\Delta \mu_o$
($\times 10^{-2} \mu_B$), and MAE (meV) for the physisorbed and chemisorbed Co$|$bilayer graphene.}
\begin{ruledtabular}
\begin{tabular}{ccccccccc}
 & $H_f$ & \emph{a} & \emph{d}$_1$ & \emph{d}$_2$ & $\mu_s^{\textrm{Co}}$ & $\mu_s^{\textrm{C}}$ & $\mu_o^{\textrm{Co}}$ & MAE \\
\hline
Physisorption & -1.12 & 2.45 & 3.29 & 3.18 & 1.96 & 0.00  & -0.81$\times$10$^3$ & --0.68\\
Chemisorption  & -2.80 & 2.50 & 1.94 & 1.58 & 1.14 & -0.04 & 0.3$\times$10$^2$ & 0.41 \\
\end{tabular}
\end{ruledtabular}
\end{table}

In addition to the magnetization reorientation,
we also address briefly the sign change of Berry curvature across the $sp^2-sp^3$ transition.
The Berry curvature is calculated using the Kubo formula\cite{guo:08,rmp:SHE}
\begin{eqnarray}
  \label{eq:1}
  \Omega_{xy}&=& \sum_{\kk,n} f(\kk,n)\Omega_{xy}(\kk,n)\\
  \Omega_{xy}(\kk,n) &=& \sum_{m}^{occ}\textrm{Im}\frac{\langle \kk,n |j_x| \kk,m\rangle\langle \kk,m | v_y| \kk,n\rangle}{\left( e_{{\bf k},n}-e_{{\bf k},m}\right)^2},
\end{eqnarray}
where $\Omega_{xy}(\kk,n)$ are
Berry curvature at $\kk$ and $n$-the band, whose summation over occupied bands and {\em k} is $\Omega_{xy}$;
$j_x=\frac{\hbar}{4}\{\sigma_z,{\bf v}\}$ is the spin-current operator, where $\sigma_z$ is Pauli matrix;
$f(\kk,n)$ is the Fermi-Dirac function; $e_{\kk,n}$ is eigenvalue of $\kk$ at $n$-the band.
In the physisorption case, the Berry curvature ($\Omega$) is 26.83 $\left(\Omega\cdot cm\right)^{-1}$
while it is --63.35 $\left(\Omega\cdot cm\right)^{-1}$ in the chemisorption.
Hence, the sign change of $\Omega$ implies the change in spin Hall effect:
spin separation occurs in opposite direction during the bond transition.
This is quite inspiring that the effect of $sp^2-sp^3$ transition can be verified experimentally
using the spin Hall effect or the inverse spin Hall effect\cite{wunderlich09,wunderlich10}
by measuring the spin current.

For more feasibility and insights,
it would be instructive to explore the other \emph{sp}$^2$ bonded layered structures.
Similar calculations are performed for the weakly deposited Co adatoms
on bottom surface fluorinated graphene bilayer.
The optimized structural and magnetic properties of the Co$|$\emph{sp}$^2$-graphene
and Co$|$\emph{sp}$^3$-carbon multilayers are tabulated in Table~\ref{tab:2}.
In terms of energetics, the formation of hybrid bond between the C-\emph{sp}$^3$ and Co-\emph{d} orbitals is
favored rather than the \emph{sp}$^2$-\emph{d} interaction,
as found in previous experiments \cite{raj13} and calculations \cite{raj13,od13}.
Notably, both the $\Delta \mu_{\textrm{o}}$ and MAE of the Co$|$\emph{sp}$^2$-graphene
and Co$|$\emph{sp}$^3$-carbon
reveal the similar features to the Co$|$BN structures.
In this sense, switching magnetization of metal adatoms by the formation of $sp^3$-$d$ hybrid bond
seems quite ubiquitous.

\subsection {IV. CONCLUSION}
In summary, our first-principles calculations predict
the reorientation of magnetization via $sp^3$-$d_{z^2}$ hybrid bond
in the Co adatoms on {\em h}-BN and graphene.
Moreover, sign change occurs not only in MAE but also in the spin Berry curvature.
Chemical functionalization of the bottom surface of \emph{h}-BN (graphene) layers leads
to the transformation into thermodynamically stable \emph{sp}$^{3}$-bonded BN (carbon) films,
which in turn develops stronger interlayer tetrahedral \emph{sp}$^{3}$-metallic \emph{d} bonds.
The present study provides a new avenue to tailor magnetization
by means of degree of the interlayer hybrid bonds in the layered materials.

\subsection {ACKNOWLEDGMENTS}
This work was supported by Incheon National University Research Grant in 20161959.
Work at Ulsan is supported 
by Creative Materials Discovery Program through the National Research Foundation of Korea(NRF) funded by the Ministry of Science, ICT and Future Planning (2015M3D1A1070465).

\end{document}